# Investigating the effects of housing instability on depression, anxiety, and mental health treatment in childhood and adolescence

Rachael Zehrung, MS[1,*], Di Hu, MS[1,*], Yawen Guo, MISM[1], Kai Zheng, PhD[1], Yunan Chen, PhD[1]
[1]University of California, Irvine, Irvine, CA, USA

**Abstract**

*Housing instability is a widespread phenomenon in the United States. In combination with other social determinants of health, housing instability affects children's overall health and development. Drawing on data from the 2022 National Survey of Children's Health, we employed multiple logistic regression models to understand how sociodemographic factors, especially housing instability, affect mental health outcomes and treatment access for youth aged 6-17 years. Our results show that youth facing housing instability have a higher likelihood of experiencing anxiety (OR: 1.42, p<0.001) and depression (OR: 1.57, p<0.001). Furthermore, youth experiencing both mental health conditions and housing instability are significantly less likely to receive mental health services in the past year, indicating the substantial barriers they face in accessing mental health care. Based on our findings, we highlight opportunities for digital mental health interventions to provide children experiencing housing instability with more accessible and consistent mental health services.*

**Introduction**

As pandemic-era protections subside and rental prices increase, housing instability in the United States has intensified, with eviction filings and cost-burdened households on the rise.[1,2] As a pervasive and multifaceted challenge, housing instability encompasses issues related to the affordability, safety, quality, and stability of housing.[3] This challenge manifests in several forms, such as overcrowding, moving frequently, missing rent or mortgage payments, and being evicted.[4] Experiencing eviction and unstable living is known to negatively affect children's well-being, increasing their risk of developmental delays, behavioral issues, and mental health conditions.[5–8] Such experiences also have long-term impacts on their life outcomes; for example, childhood housing instability is associated with a higher risk of depression and anxiety in adulthood.[9] Moreover, housing instability disproportionately burdens youth with marginalized identities (e.g., BIPOC and LGBTQ+), who are already vulnerable to mental health challenges due to other compounding social inequalities (e.g., discrimination).[10]

Prior work has found that housing instability amplifies both physical and mental health disparities, demonstrating its role as a critical social determinant of health.[11,12] Social determinants of health (SDoH), defined as "the conditions in which people are born, grow, work, live, and age, and the wider set of forces and systems shaping the conditions of daily life,"[13] include factors such as economic stability, education access and quality, neighborhood and built environment, social and community context, and healthcare access and quality.[14] A growing body of research explores the complex interplay among SDoH, emphasizing the influence of upstream social determinants (e.g., socioeconomic resources) on downstream social determinants (e.g., health behaviors).[15,16] For instance, housing instability is linked to postponed medical care and increased emergency department visits for children in low-income families.[17] These findings underscore the importance and necessity of understanding SDoH to address health inequity and improve child health outcomes.

Families facing housing instability may also encounter barriers such as lacking a primary care provider, health insurance, and transportation tools, which can impede children from accessing necessary health care.[18,19] Despite the evident effects of housing instability on children's mental health, little work has examined how housing instability influences children's usage of mental health services. This study bridges this gap by examining how various SDoH, especially housing instability, affect children's mental health outcomes and their access to mental health care. In light of the challenges of accessing care faced by youth experiencing housing instability, technology-based mental health interventions show promise, though existing work tends to focus on older adolescents and young adults.[20,21] Building on the risk and protective factors identified in our study for children experiencing housing instability, we outline the potential opportunities for digital mental health interventions to support the delivery of mental health services to this population.

---

[*] The authors contributed equally to this work

With the goal of identifying opportunities for digital mental health interventions, this study investigates the following research questions:
1. How do housing instability and other sociodemographic factors influence the occurrence of anxiety and depression among youth aged 6-17 years old?
2. How does housing instability affect mental health treatment usage among youth aged 6-17 years old with mental health conditions?

We hypothesized that housing instability increases the likelihood of mental health conditions in children while significantly hindering their access to mental health treatment.

**Methods**
This study first investigated a range of social and demographic factors that influence youth mental health and then examined the effect of housing instability on youth with mental health conditions. In this section, we describe our data source, selection of independent and dependent variables, and data analysis approach.

*Data Source*. The data used in this study come from the 2022 National Survey of Children's Health (NSCH), which is a nationally representative survey on the physical and emotional health of children (aged 0-17) in the United States. The survey is directed by the Health Resources and Service Administration's Maternal and Child Health Bureau (HRSA MCHB) and conducted by the U.S. Census Bureau on an annual basis. The NSCH uses a national sample of addresses and administers a screening questionnaire to identify households with children, followed by an age-based topical questionnaire for eligible households. The 2022 NSCH used a sample of 360,000 addresses with a weighted overall response rate of 39.1%. Of the 122,000 completed screener questionnaires, 67,269 households were eligible for a topiccal questionnaire follow-up, which 54,103 households completed. Our analysis considered children aged 6-17, resulting in a dataset with 34,362 households before data cleaning. More details on the survey methodology can be found in the U.S. Census Bureau's methodology report.[22]

*Dependent Variables.* The occurrence of anxiety/depression and use of mental health treatment were defined as dependent variables (Table 1). We categorized a child as having anxiety or depression only if the respondent indicated the child currently has the condition because we aimed to examine relationships between children's current living situations and mental health. We re-coded the question asking about mental health treatment reception (K2Q22_R) to be binary (i.e., "yes" or "no" to receiving treatment).

**Table 1.** Survey questions used to assess the occurrence of depression/ anxiety and use of mental health treatment

| Q_ID | Question | Options |
|---|---|---|
| K2Q32A | Has a doctor or other health care provider EVER told you that this child has Depression? | 1) Yes<br>2) No |
| K2Q32B | If yes, does this child CURRENTLY have the condition? | 1) Yes<br>2) No |
| K2Q33A | Has a doctor or other health care provider EVER told you that this child has Anxiety Problems? | 1) Yes<br>2) No |
| K2Q33B | If yes, does this child CURRENTLY have the condition? | 1) Yes<br>2) No |
| K4Q22_R | DURING THE PAST 12 MONTHS, has this child received any treatment or counseling from a mental health professional?<br><br>Mental health professionals include psychiatrists, psychologists, psychiatric nurses, and clinical social workers. | 1) Yes<br>2) No, but this child needed to see a mental health professional<br>3) No, this child did not need to see a mental health professional |

*Independent Variables*. Informed by prior work on housing instability and the SDoH framework, we selected social and demographic factors to understand their associations with youth mental health. For demographic information, we considered factors such as children's age, gender, race, body mass index (BMI), and family structure. Independent variables in our final analyses were selected based on theoretical relevance to our research objectives and a multicollinearity test (Variance Inflation Factor).

Housing Instability: Based on definitions of housing instability,[23,24] we used a combination of three survey questions to assess families' housing situations (Table 2). Given that the threat of eviction and eviction itself are significant indicators of housing instability,[25] we identified respondents as housing unstable if they were always, usually, or sometimes worried about eviction (HOMEEVIC). In line with prior research on housing instability, we combined multiple moves (PLACESLIVED) and being behind on rent or mortgage payments (MISSMORTGAGE) in the past year as one indicator.[26] Prior work has found that the number of moves is not sensitive enough as a single-item indicator to determine housing instability, as moving could be voluntary and positive.[23] Similarly, missing one rent or mortgage payment might not reflect longitudinal patterns of housing instability.[27] An exploratory analysis confirmed that the combination of PLACESLIVED and MISSMORTGAGE was a more reliable indicator of housing instability than either one individually.

**Table 2**. Survey questions used to assess housing instability

| Q_ID | Question | Options |
|---|---|---|
| HOMEEVIC | DURING THE PAST 12 MONTHS, how often were you worried or stressed about being evicted, foreclosed on, or having your housing condemned? | 1) Always<br>2) Usually<br>3) Sometimes<br>4) Rarely<br>5) Never |
| MISSMORTGAGE | DURING THE PAST 12 MONTHS, was there a time when you were not able to pay the mortgage or rent on time? | 1) Yes<br>2) No<br>3) Don't Know |
| PLACESLIVED | DURING THE PAST 12 MONTHS, how many places has this child lived? | 1) 0-2 places<br>2) 3 or more places |

Neighborhood Environment: As SDoH, the neighborhoods that people live in and the interactions that they have within their communities have an impact on their overall health and well-being.[14,28] Social support, for example, can buffer the negative effects imposed by other risks to health.[29] We included questions that asked respondents about their subjective views on their community (e.g., school safety) and the presence of certain features in their neighborhoods (e.g., a recreation center). By including these factors, we aimed to develop a more complete understanding of respondents' living situations and which factors contribute to children's mental health.

**Table 3**. Survey questions used to define neighborhood characteristics

| Q_ID | Question | Options |
|---|---|---|
| K10Q13 | In your neighborhood, is/are there:<br>A recreation center, community center, or boys' and girls' club? | 1) Yes<br>2) No |
| K10Q23 | In your neighborhood, is/are there:<br>Vandalism such as broken windows or graffiti? | 1) Yes<br>2) No |
| K10Q14 | To what extent do you agree with these statements about your neighborhood or community?<br><br>This child is safe at school | 1) Definitely agree<br>2) Somewhat agree<br>3) Somewhat disagree<br>4) Definitely disagree |

Social Context: Children's participation in extracurricular activities is associated with positive psychological and behavioral outcomes, such as higher self-esteem and lower rates of depression.[30,31] Extracurricular activities can be particularly beneficial for marginalized youth, who tend to participate less but experience greater benefits.[32,33] Recognizing the impact of extracurricular activities on children's health, we included relevant survey questions about extracurricular participation to understand their effect on the mental health of children experiencing housing instability (Table 4).

**Table 4.** Survey questions used to account for children's participation in extracurricular activities

| Q_ID | Question: DURING THE PAST 12 MONTHS, did this child participate in: | Options |
|---|---|---|
| K7Q30 | A sports team or did they take sports lessons after school or on weekends? | 1) Yes 2) No |
| K7Q32 | Any other organized activities or lessons, such as music, dance, language, or other arts? | 1) Yes 2) No |
| K7Q37 | Any type of community service or volunteer work at school, place of worship, or in the community? | 1) Yes 2) No |

*Data Analysis*. NSCH survey data were analyzed using R software (version 4.3.3). Prior to conducting any statistical analyses, we performed data cleaning and preprocessing which included re-coding certain questions to define our dependent and independent variables (as described above) and excluding cases with missing values. Data from 30,364 households ended up being included in our final analysis. To explore the influence of sociodemographic factors on the occurrence of anxiety and depression (RQ1), we employed multiple logistic regression models for each condition. We hypothesized that sociodemographic factors play significant roles in explaining variations in mental health outcomes among youth aged 6-17 years. To examine how housing instability affects mental health treatment usage, particularly among youth with mental health conditions (RQ2), we first constructed a multiple logistic regression model with 12-month mental health treatment usage as the dependent variable, controlling for all other sociodemographic factors. We then introduced interaction terms between the occurrence of mental health conditions (anxiety and depression, respectively) and housing instability to specifically examine the impact of housing instability on mental health treatment usage among youth with mental conditions. This focus is driven by our interest in the subgroup most affected by these factors, rather than in youth without mental health issues. We hypothesize that youth with mental health conditions, although in need, are less likely to receive mental health treatment services if they are experiencing housing instability.

**Results**

*Descriptive Statistics*. In the final analytical sample of 30,364 children, 51.9% were male. In terms of race, 77.3% were White, 6.7% were Black or African American, and the remaining 16% were identified as Other. The mean age of children was 12 (Standard Deviation: 3.5). Within this population, 7.1% were identified as experiencing housing instability. Furthermore, 16.4% and 7.3% were reported as currently experiencing anxiety and depression, respectively. In the 12 months prior to the survey, 17.9% had received professional mental health treatment. Based on this sample, we conducted regression analysis to answer our research questions.

*RQ1: How do housing instability and other sociodemographic factors influence the occurrence of anxiety and depression among youth aged 6-17 years old?* In this section, we present the results from the logistic regression models (Table 5) that used the occurrence of anxiety and depression as dependent variables.

<u>Demographic Factors:</u> We found that sex, age, race, and family structure were significant predictors of anxiety and depression. Female children had twice the odds of having depression ($p<0.001$) and 57% higher odds of having anxiety ($p<0.001$) compared to male children. As a child's age increased by one year, their odds of experiencing anxiety and depression increased by 12% and 29% ($p<0.001$), respectively. Our analysis showed that Black children and other children of color were significantly less likely to have depression and anxiety.

<u>Housing Instability:</u> We found that housing instability had a significant effect on the occurrence of anxiety and depression. Controlling for other sociodemographic factors, children experiencing housing instability were more likely to have anxiety (OR: 1.42, $p<0.001$) and depression (OR: 1.57, $p<0.001$). Driven by our particular interest in the housing unstable population, we conducted a follow-up subgroup analysis after reviewing these results from the general models. Comparing children in stable housing and children in unstable housing, we found that social and demographic factors affected children at similar rates regardless of housing status, with the exception of family structure. Children who lived with relatives other than parents or grandparents in unstable housing were significantly more likely to have depression (OR: 14.23, p<0.001) and anxiety (OR: 4.20, p<0.001) than children who lived with other relatives in stable housing.

**Table 5.** Model results with "current_anxiety" and "current_depression" as dependent variables.

| Variable | Current_Anxiety | | Current_Depression | |
|---|---|---|---|---|
| | Odds.Ratio | P-value / 95% CI | Odds.Ratio | P-value / 95% CI |
| Sex (Female) | 1.57 | *** (1.47 - 1.68) | 2.03 | *** (1.84 - 2.24) |
| Age | 1.12 | *** (1.11 - 1.13) | 1.29 | *** (1.27 - 1.31) |
| Race (White) | Reference | | | |
| Race (Black) | 0.33 | *** (0.28 - 0.39) | 0.36 | *** (0.28 - 0.44) |
| Race (Other) | 0.57 | *** (0.52 - 0.63) | 0.74 | *** (0.64 - 0.84) |
| BMI (<5th percentile) | Reference | | | |
| BMI (5-85th percentile) | 0.94 | (0.83 - 1.06) | 0.93 | (0.77 - 1.14) |
| BMI (85-95th percentile) | 1.03 | (0.89 - 1.18) | 1.18 | (0.95 - 1.47) |
| BMI (>=95th percentile) | 1.15 | * (1 - 1.32) | 1.58 | *** (1.28 - 1.96) |
| Family_structure (Two biological/adoptive parents, married) | Reference | | | |
| Family_structure (Two biological/adoptive parents, non-married) | 0.91 | (0.75 - 1.1) | 1.05 | (0.78 - 1.39) |
| Family_structure (Two parents, married, at least one not biological/adoptive) | 1.42 | *** (1.26 - 1.61) | 1.87 | *** (1.58 - 2.2) |
| Family_structure (Two parents, non-married, at least one not biological/adoptive) | 1.52 | *** (1.23 - 1.88) | 2.61 | *** (1.99 - 3.4) |
| Family_structure (Single mother) | 1.39 | *** (1.27 - 1.51) | 1.82 | *** (1.61 - 2.05) |
| Family_structure (Single father) | 0.70 | *** (0.6 - 0.82) | 0.94 | (0.74 - 1.17) |
| Family_structure (Grandparent household) | 1.68 | *** (1.41 - 2) | 2.88 | *** (2.3 - 3.59) |
| Family_structure (Other relation) | 1.89 | *** (1.43 - 2.49) | 3.67 | *** (2.66 - 5.02) |
| Sports Team or Lessons (No) | 1.58 | *** (1.48 - 1.69) | 1.86 | *** (1.68 - 2.05) |
| Other Organized Activities or Lessons (No) | 0.95 | (0.89 - 1.02) | 1.13 | * (1.02 - 1.25) |
| Community Service or Volunteer Work (No) | 1.23 | *** (1.14 - 1.32) | 1.29 | *** (1.16 - 1.43) |
| Neighborhood_Vandalism (No) | 0.81 | *** (0.72 - 0.91) | 0.81 | ** (0.68 - 0.95) |
| Neighborhood_Recreation Center (No) | 1.03 | (0.97 - 1.1) | 1.18 | *** (1.07 - 1.3) |
| Safe_at_School (Definitely Agree) | Reference | | | |
| Safe_at_School (Somewhat Agree) | 1.51 | *** (1.41 - 1.61) | 1.70 | *** (1.54 - 1.87) |
| Safe_at_School (Somewhat Disagree) | 2.36 | *** (2.03 - 2.74) | 3.09 | *** (2.56 - 3.72) |
| Safe_at_School (Definitely Disagree) | 4.24 | *** (3.34 - 5.38) | 5.61 | *** (4.24 - 7.39) |
| Housing_Instability (Yes) | 1.42 | *** (1.27 - 1.59) | 1.57 | *** (1.36 - 1.82) |

p-value <0.001 ***, <0.01 **, <0.05 *

<u>Social Factors:</u> Children in neighborhoods with vandalism were more likely to have anxiety (p<0.001) and depression (p<0.01). Likewise, children in neighborhoods without a recreation center were more likely to have depression (p<0.001), though there was no effect on anxiety. As parental perceptions of children's safety at school decreased, the odds of children having anxiety and depression increased. For example, when parents felt that their children were definitely not safe at school, children had 4.24 times the odds of having anxiety (p<0.001) and 5.61 times the odds of having depression (p<0.001). Taken together, these results suggest that neighborhood and community characteristics have a significant effect on children's mental health, particularly when those characteristics relate to perceptions of safety. Further, we found that children who did not participate in sports were more likely to have anxiety (OR: 1.58, p<0.001) and depression (OR: 1.86, p<0.001). Similarly, children who did not participate in community service or volunteer work were more likely to have anxiety (OR: 1.23, p<0.001) and depression (OR: 1.29, p<0.001). Participation in other organized activities or lessons had no significant effect on anxiety and a smaller effect on depression (OR: 1.13, p<0.05).

*RQ2: How does housing instability affect mental health treatment usage among youth aged 6-17 years old with mental health conditions?* Despite the increased likelihood of mental health conditions among the housing unstable population, the effect of depression on mental health service use was reduced by 38% (p<0.01) for children also experiencing housing instability. Similarly, the effect of anxiety on mental health service use was reduced by 39% (p<0.001) for these children. The role of housing stability is further evident in that children with depression have 22 times the odds (p<0.001) of receiving mental health treatment if they had stable housing. Likewise, children with anxiety have 17 times the odds (p<0.001) of receiving treatment if they had stable housing.

**Table 6.** Model results with "Mental_Health_Treatment" as the dependent variable, controlling all other sociodemographic factors in the previous models.

Table 6.1, adding an interaction term between "Current_Anxiety" and "Housing_Instability."

| | | P-value |
|---|---|---|
| **Interaction Terms** | **Odds.Ratio** | **95% CI** |

| | | *** |
|---|---|---|
| Current_Anxiety (Yes) | 17.4 | (16.03 - 18.79) |
| Housing_Instability (Yes) | 1.01 | (0.85 - 1.2) |
| Current_Anxiety (Yes) : Housing_Instability (Yes) | 0.61 | *** (0.47 - 0.78) |

p-value <0.001 ***, <0.01 **, <0.05 *

Table 6.2, adding an interaction term between "Current_Depression" and "Housing_Instability."

| Interaction Terms | Odds.Ratio | P-value 95% CI |
|---|---|---|
| Current_Depression (Yes) | 22.33 | *** (19.73 - 25.33) |
| Housing_Instability (Yes) | 0.91 | (0.78 - 1.05) |
| Current_Depression (Yes) : Housing_Instability (Yes) | 0.62 | ** (0.46 - 0.85) |

p-value <0.001 ***, <0.01 **, <0.05 *

**Discussion**

This study investigated the effects of housing instability and other SDoH on children's mental health outcomes and access to mental health treatment. Echoing prior work, our analysis revealed disparities in mental health outcomes based on demographic and social factors. Specifically, we found that children experiencing housing instability were significantly more likely to have anxiety and depression, yet these children with mental health conditions were less likely to receive mental health treatment compared to their counterparts in stable housing. Families experiencing housing instability encounter financial, structural, and cognitive barriers to healthcare access and service utilization.[34] These barriers include limited health insurance, lack of a primary care provider, and difficulty navigating the healthcare system, which disrupt the continuity of care for children. To overcome these barriers to care and address mental health disparities among marginalized populations, prior work has highlighted the potential for digital mental health (DMH) solutions due to their accessibility, scalability, and cost effectiveness.[35] Digital mental health technologies include web-based platforms, mobile applications, wearable devices, and other tools used to monitor health, share health-related information, and deliver care. Although youth cannot change their housing situations or actively seek professional treatment on their own, DMH technologies can be used to bridge the gap in treatment access by providing youth with free or low-cost services that they can access from anywhere. Despite extensive research on DMH for both children and individuals with unstable housing, little work has examined the intersection of these populations. Our work contributes to the growing call to consider housing stability as a fundamental aspect of children's health. Based on our findings, we discuss considerations for DMH interventions that target youth experiencing housing instability.

*Considering Children of Different Ages:* In designing DMH technologies for children, it is imperative to consider the role of both demographic factors (e.g., age) and technical factors (e.g., device ownership). Existing DMH interventions for youth tend to focus on adolescents and young adults, while little work engages with younger children under the age of 10.[36] However, symptoms of anxiety and depression can onset early in childhood,[37,38] and our results confirmed that children face an increased mental health risk as they age. That said, delivering DMH interventions to younger children presents challenges. While smartphone ownership rates are high among adolescents regardless of socioeconomic background, only a minority of children under the age of 11 have their own smartphones.[39,40] Due to low rates of device ownership among younger children, we highlight the potential of collaborative DMH interventions that involve caregivers in scaffolding children's social emotional learning.[41] Prior work has found that DMH interventions for youth tend to be more effective when they involve regular interactions with a therapist, peer, or parent, in comparison to fully automated or self-administered interventions.[42] Given that youth experiencing housing instability might have less access to therapists or parents (e.g., if they live with other relatives), future research should explore the role of other caregivers in facilitating DMH. Though DMH technologies cannot replace professional treatment, they can bridge the gap for children with less access to mental health services. For youth with their own devices, DMH interventions can be individual or collaborative.

*Considering Social Connection and Context*: Research suggests that youth value social connection when using digital health technologies, particularly with peers experiencing similar challenges.[43] In particular, adolescents perceive peer communication and support as essential aids in managing their mental health.[44] For youth experiencing housing instability, higher levels of social connectedness have been associated with resilience and improved mental health.[45] This perspective aligns with our findings, which demonstrated the positive effect of social involvement (e.g., sports activities and community service) on children's mental health. While children experiencing housing instability face barriers to participating in group and community activities in-person, often as a result of frequent moves, the online connectedness facilitated by DMH technologies may offer them opportunities to tackle this issue.[46] Furthermore, DMH technologies should consider the social context of youth experiencing housing instability; for example, living with relatives other than parents or grandparents might reduce children's sense of privacy and increase their reluctance to openly use DMH tools in front of or with help from others. Future work should explore the needs and preferences of youth experiencing housing instability by actively engaging them in the co-design process to ensure the acceptability of the final design.

*Limitations*: Our analysis showed that BIPOC youth exhibited lower rates of depression and anxiety; however, these findings might not accurately represent the mental health landscape due to racial disparities within the survey population, as over 75% of the population was White. Further, BIPOC youth may be underdiagnosed (or misdiagnosed) and undertreated for depression and anxiety.[47] Our model for predicting mental health service in the past 12 months supports this claim by verifying that Black children and children of color were less likely to have received any professional mental health service. Our analysis was further limited by the questions asked in the survey. For example, we were unable to evaluate reasons for why children did not receive mental health treatment. Future work might consider integrating quantitative and qualitative approaches to gain more insight into the challenges faced by families experiencing housing instability as they navigate their children's mental health. We also emphasize that while DMH interventions often focus on individual behaviors, it is crucial for policymakers to address upstream SDoH (e.g., housing policies) to combat fundamental causes of health inequalities.

**Conclusion**
Our study underscores the significant impact of housing instability on children's mental health outcomes and access to mental health services. We found that youth facing housing instability are at a heightened risk of experiencing anxiety and depression, highlighting the need for targeted interventions to address these disparities. Furthermore, our findings revealed significant gaps in mental health treatment access for children experiencing both housing instability and mental health conditions. This study contributes to an understanding of the complex interplay between SDoH and children's mental health, emphasizing the importance of addressing housing instability as a public health concern. Based on our findings, we discussed the design of age-appropriate and socially supportive DMH interventions tailored to the unique needs and challenges of children experiencing housing instability.